\documentclass[a4paper,10pt]{article}

\usepackage{amssymb,amsmath,amscd}

\newtheorem{definition}{Definition}[section]
\newtheorem{lemma}[definition]{Lemma}
\newtheorem{proposition}[definition]{Proposition}
\newtheorem{theorem}[definition]{Theorem}

\newenvironment{proof*}{\smallskip\par\noindent\emph{Proof. }
 \ignorespaces}{\hfill$\Box$\smallskip\par\ignorespaces}
\newenvironment{proofsketch*}{\smallskip\par\noindent
 \emph{Sketch of proof. }\ignorespaces}
 {\hfill$\oslash$\smallskip\par\ignorespaces}

\newcommand{\Ws}{\ensuremath{:\!\phi^2\:\!\!\!:\!}}

\title{Local vs.~global temperature\\ under a positive curvature condition}
\author{Ko Sanders\thanks{jacobus.sanders@dcu.ie}\\
Dublin City University\\
School of Mathematical Sciences\\
Glasnevin\\
Dublin 9, Ireland}
\date{21 May 2017}

\begin{document}

\maketitle

\begin{abstract}
For a massless free scalar field in a globally hyperbolic space-time we compare the global temperature $T=\beta^{-1}$, defined for the $\beta$-KMS states $\omega^{(\beta)}$, with the local temperature $T_{\omega}(x)$ introduced by Buchholz and Schlemmer. We prove the following claims: (1) Whenever $T_{\omega^{(\beta)}}(x)$ is defined, it is a continuous, monotonically decreasing function of $\beta$ at every point $x$. (2) $T_{\omega}(x)$ is defined when $M$ is ultra-static with compact Cauchy surface and non-trivial scalar curvature $R\ge0$, $\omega$ is stationary and a few other assumptions are satisfied. Our proof of (2) relies on the positive mass theorem. We discuss the necessity of its assumptions, providing counter-examples in an ultra-static space-time with non-compact Cauchy surface and $R<0$ somewhere. Our results suggest that under suitable circumstances (in particular in the absence of acceleration, rotation and violations of the weak energy condition in the background space-time) both notions of temperature provide qualitatively similar information, and hence the Wick square can be used as a local thermometer.
\end{abstract}

\section{Introduction to global and local temperature}\label{Sec_Introduction}

Consider a connected globally hyperbolic space-time $M$ with signature $(-+++)$, whose metric $g_{ab}$ is related by Einstein's equation to classical matter with stress tensor\footnote{We will work in units where $c=\hbar=G=k_B=1$.}
\begin{equation}\label{Eq_Einstein}
T^{\mathrm{cl}}_{ab}=\frac{1}{8\pi}G_{ab}=\frac{1}{8\pi}\left(R_{ab}-\frac12Rg_{ab}\right),
\end{equation}
where $R_{ab}$ is the Ricci curvature and $R$ the scalar curvature of the metric. On this space-time we consider a scalar quantum field $\phi$ satisfying the Klein-Gordon equation with a smooth external potential $V$:
\begin{equation}\label{Eq_KG}
(-\Box_g+V)\phi=0,
\end{equation}
where $\Box_g$ denotes the covariant d'Alembert operator. We will often take $V=m^2+\xi R$ with a mass $m\ge0$ and scalar curvature coupling $\xi\in\mathbb{R}$. The general theory of a wave equation like (\ref{Eq_KG}) is well understood (see e.g.~\cite{Khavkine+2014_2}). Note in particular that we treat $\phi$ as a test-field, keeping the metric fixed, so in general the semi-classical Einstein equation is violated,
\begin{equation}\label{Eq_SCEinstein}
\omega(T^{\mathrm{ren}}_{ab}(\phi))+T^{\mathrm{cl}}_{ab}\not=\frac{1}{8\pi}G_{ab},
\end{equation}
and the expectation value of the renormalised stress tensor $T^{\mathrm{ren}}_{ab}(\phi)$ of the quantum field $\phi$ in the state $\omega$ may be interpreted as the amount of stress, energy and momentum that are injected into (or extracted from) the quantum field by keeping the external metric fixed.

To define global thermal equilibrium states we assume that $M$ is stationary, so the theory has a preferred time flow. A state $\omega$ which satisfies the $\beta$-KMS condition \cite{Kubo1957,Martin+1959,Bratteli+1997} w.r.t.~this time flow for some $\beta\in(0,\infty]$ is a global thermal equilibrium state at temperature (in natural units)
\begin{equation}\label{Eqn_GT}
T=\beta^{-1}.
\end{equation}
We will denote such states by $\omega^{(\beta)}$, where the case $\beta=\infty$ corresponds to the ground state. The existence of $\beta$-KMS states for all $\beta>0$ can be proven under suitable circumstances (e.g.~when $V$ is stationary and $V>0$ everywhere, cf.~\cite{Sanders2013} and section \ref{SSec_Wick} below). The interpretation of $\beta$-KMS states as global thermal equilibrium states may be motivated by arguments from quantum statistical mechanics (cf.~\cite{Bratteli+1997} and references therein).

A $\beta$-KMS state $\omega$ is necessarily stationary, i.e.~it is invariant under the time flow. This is an inherently global property, because essentially any two regions of space can communicate with each other, given enough time. Hence a local perturbation of a $\beta$-KMS state cannot be expected to be a stationary state again. Consequently, the temperature in (\ref{Eqn_GT}) is also an inherently global property. This raises the question how it relates to the temperature as measured locally by an observer. Especially in curved space-times one expects several complications: (i) stationary observers may be accelerating or rotating, which leads to apparent forces on the systems they observe; (ii) keeping the metric fixed leads to a position-dependent injection of energy into the quantum field. The first of these complications has been extensively debated in the context of the Unruh effect \cite{Unruh1976,Earman2011,Crispino+2008,Buchholz+2013_3,Buchholz+2014_2}, so let us focus on the second. Suppose that the classical background metric is stationary, but violates the weak energy condition. In this case one might expect that the field $\phi$, in a $\beta$-KMS state, can transfer energy (and entropy) to the classical matter, via the metric. When the back-reaction of the metric is not taken into account, this process can continue indefinitely, casting doubt on the equilibrium character of the state. In particular, such a constellation would make a thermodynamical interpretation of the temperature $T$ of a $\beta$-KMS state problematic.

In an attempt to bypass such problems with the global temperature one may consider instead a local notion of temperature. One such notion was proposed by Buchholz, Ojima and Roos \cite{Buchholz+2002_2} in their framework for local thermal equilibrium, and it was extended to a generally covariant setting by Buchholz and Schlemmer \cite{Buchholz+2007} (see also \cite{Solveen2010,Gransee+2015}). It is the purpose of this paper to compare this local notion of temperature with the global notion.

To define the local notion of temperature we may return to the setting of a general globally hyperbolic space-time $M$. We will consider states $\omega$ for the quantum theory which are Hadamard \cite{Kay+1991,Radzikowski1996}. This includes all $\beta$-KMS states (see e.g.~\cite{Strohmaier+2002}). In general, Hadamard states are characterised by a condition on their two-point distribution $\omega_2(x,y)$. For each $k\in\mathbb{N}$ there is a distribution $H^{(k)}_2$, defined on a neighbourhood of the diagonal in $M\times M$ and constructed in a local and covariant way out of the metric and the Klein-Gordon operator $-\Box_g+V$, such that $\omega_2$ is Hadamard if and only if $\omega_2-H_2^{(k)}$ is a $C^k$ function for all $k\in\mathbb{N}$. Given the distributions $H^{(k)}_2$ for all space-times $M$, one can define a generally covariant renormalisation scheme for the Wick square $\Ws$ and other objects \cite{Brunetti+1996,Hollands+2001}. By construction, the expectation value
\begin{equation}\label{Eqn_Wicksquare}
\omega(\Ws(x)):=\lim_{y\to x}\omega_2(x,y)-H_2^{(k)}(x,y),\qquad k\ge 0,
\end{equation}
is then a smooth function of $x\in M$, which depends on $\omega_2$, but not on $k$.

The distributions $H_2^{(k)}$ are not unique \cite{Hollands+2001}, but the renormalisation freedom in $\Ws\ $ reduces to multiples of the mass square and the scalar curvature. Any alternative renormalisation scheme, indicated by a prime, takes the form
\begin{equation}\label{Eqn_freedom}
\Ws'=\ \Ws+c_1R+c_2m^2
\end{equation}
for constants $c_1,c_2\in\mathbb{R}$.

Following \cite{Buchholz+2007} we define the local temperature at a point $x$ of a massless free field ($V=\xi R$) in a Hadamard state $\omega$ by\footnote{\cite{Lynch+2016} recently proposed a generalised instantaneous temperature, which differs from equation (\ref{Eqn_LT}) by local modifications depending on the curvature and the observer's acceleration. The derivation of this formula uses Unruh-DeWitt detectors with a large energy gap and a short interaction interval. We refer to \cite{Fewster+2016} for critical notes on the limitations of this kind of derivation and we note that we will mostly focus on situations where the two formulae agree.}
\begin{equation}\label{Eqn_LT}
T_{\omega}(x):=\sqrt{12\omega(\Ws(x))}
\end{equation}
whenever the expectation value of $\Ws(x)$ is non-negative. When the expectation value is negative, then the state is not in local thermal equilibrium at the point $x$ and the local temperature is not defined.

Two comments on this definition of local temperature are in order. First, for $\beta$-KMS states of a massless free scalar field in Minkowski space, the definition above yields
\begin{equation}\label{Eqn_M0m0}
T_{\omega}(x)\equiv\beta^{-1}=T,
\end{equation}
i.e.~the local temperature at any point $x$ equals the global temperature. This motivates the use of the Wick square $\Ws(x)$ as a local thermometer. Because the renormalisation scheme is generally covariant, \cite{Buchholz+2007} suggests that the same formula should work also in curved space-times, at least for massless free scalar fields (and perhaps imposing conformal coupling, $\xi=\frac16$). For massive fields in Minkowski space, the relation between $\beta$ and $\omega^{(\beta)}(\Ws(x))$ is different, and equation (\ref{Eqn_LT}) may have to be modified accordingly.

Secondly, the reader will notice that the value of the local temperature, and even its existence, depends on the renormalisation scheme chosen (see \cite{Buchholz+2007} for related remarks). Because our main interest is not in the renormalisation ambiguities, we will avoid them when we investigate the existence of a local temperature by restricting attention to points where $R(x)=0$.

At face value the definition (\ref{Eqn_LT}) may seem rather arbitrary, because there could be many local observables in Minkowski space from which the temperature $T$ can be recovered. Moreover, as we will discuss in section \ref{Sec_noT}, there are many situations in which $T_{\omega}(x)$ is ill-defined because $\omega(\Ws(x))$ is negative. This includes situations where $\omega$ is not stationary (cf.~quantum inequalities \cite{Fewster2005,Solveen2010}), or when the stationary observers are accelerated (e.g.~the Fulling vacuum in Rindler space-time \cite{Buchholz+2013_3}). We will identify an additional cause for $T_{\omega}(x)$ to become ill-defined: a violation of the weak energy condition. As an example we will construct ultra-static space-times with $R<0$ somewhere, whose ground states have no well-defined local temperature in a flat region of the space-time.

Our main purpose, however, is to substantiate the claim that the local and global notions of temperature do provide qualitatively similar information, and hence the Wick square can be used as a local thermometer, as long as certain physically relevant restrictions are imposed (see also \cite{Solveen2012,Schlemmer+2008}). In section \ref{Sec_monotonic} we show that the expectation values $\omega^{(\beta)}(\Ws(x))$ are continuously and monotonically decreasing in $\beta$ at any point $x$, and the minimum over all stationary sates is achieved by the ground state (if it exists). In section \ref{Sec_existence} we show moreover that $T_{\omega^{\beta}}(x)$ is well-defined for all stationary states in an ultra-static space-time when the scalar curvature satisfies $R\ge0$ and when a few technical conditions on the space-time, the point $x$, and the scalar curvature coupling of the field are met. The proof adapts an argument of Schoen \cite{Schoen1984}, which relies on the positive mass theorem \cite{Schoen+1979}. Note that $R\ge0$ can only be violated if the stress tensor $T^{\mathrm{cl}}_{ab}$ of the classical background matter violates the weak, strong and dominant energy conditions, which are all equivalent to $R_{ab}\ge0$ in the static case. The questions whether the technical conditions can be weakened, and whether the result can be generalised, will be raised in the final section \ref{Sec_discussion}.

\section{Monotonicity of the local temperature}\label{Sec_monotonic}

Let us consider a stationary space-time $M$ with complete future pointing time-like Killing vector field $\xi^a$. We also assume that $V$ is stationary, i.e.~that $\xi^a\nabla_aV=0$.

For a classical solution $\varphi$ to the Klein-Gordon equation (\ref{Eq_KG}) one defines the stress-energy-momentum tensor
\begin{equation}\label{Eqn_T}
T_{ab}(\varphi):=\nabla_{(a}\bar{\varphi}\nabla_{b)}\varphi-\frac12g_{ab}(\nabla^c\bar{\varphi}\nabla_c\varphi+V|\varphi|^2)
\end{equation}
and when the integral converges absolutely one defines the energy by
\begin{equation}\label{Eqn_energy}
\mathcal{E}(\varphi):=\int_{\Sigma}n^a\xi^bT_{ab}(\varphi),
\end{equation}
where $n^a$ is the future pointing normal vector field to the Cauchy surface $\Sigma$ and we integrate w.r.t.~the invariant volume form of the induced metric on $\Sigma$. $\mathcal{E}(\varphi)$ is independent of the choice of $\Sigma$, because $\nabla^a\xi^bT_{ab}(\varphi)=0$.

The Klein-Gordon equation (\ref{Eq_KG}) admits unique advanced ($-$) and retarded ($+$) fundamental solutions $E^{\pm}:C_0^{\infty}(M)\to C^{\infty}(M)$, and we write $E:=E^--E^+$. For any test-function $f\in C_0^{\infty}(M)$, $E(f)$ is a space-like compact solution to (\ref{Eq_KG}) which has finite energy. When $V\ge0$ and either $V$ is non-trivial or $\Sigma$ has infinite volume, then $\sqrt{\mathcal{E}(\varphi)}$ defines a norm on the linear space of solutions $E(f)$. Completing this space in this norm yields the Hilbert space $\mathcal{H}_{cl}$ of classical finite energy solutions. The time flow determined by $\xi^a$ preserves the space of solutions and their energy and it determines a one-parameter unitary group on $\mathcal{H}_{cl}$, which is generated by a Hamiltonian $H$.

When $H$ is invertible and the range of $E$ is in the domain of $|H|^{-1}$, then we can formulate the two-point distributions of $\beta$-KMS states in terms of these classical structures as follows (cf.~\cite{Sanders2013}). We introduce the two-point distributions
\begin{equation}\label{Eqn_KMS}
\omega^{(\beta)}_2(\bar{f},f'):=2\langle E(f), H^{-1}(e^{\beta H}-1)^{-1}E(f')\rangle,
\end{equation}
for all $\beta>0$, where the limit $\beta\to\infty$ is given by
\begin{equation}\label{Eqn_ground}
\omega^{(\infty)}_2(\bar{f},f'):=2\langle E(f), P_-|H|^{-1}E(f')\rangle,
\end{equation}
with $P_-$ the spectral projection of $H$ onto $(-\infty,0)$. These two-point distributions give rise to quasi-free states which satisfy the $\beta$-KMS condition.\footnote{In principle there may be other $\beta$-KMS states \cite{Sanders2013}, but these are usually excluded, e.g.~because of their bad behaviour at spatial infinity, or because they lack the clustering property \cite{Haag}.}

\begin{proposition}\label{Prop_monotonic}
At any point $x\in M$, the function $\beta\mapsto\omega^{(\beta)}_2(\Ws(x))$ is continuous and monotonically decreasing in $\beta\in(0,\infty]$.
\end{proposition}
\begin{proof*}
For every $0<\beta<\infty$ we define the function $F_{\beta}(k)=(e^{\beta k}-1)^{-1}$ on $k>0$, which takes only positive values. We note that for any $x\in M$
\begin{eqnarray}
\omega^{(\beta)}_2(\Ws(x))-\omega^{(\infty)}_2(\Ws(x))&=&\lim_{n\to\infty}(\omega^{(\beta)}_2-\omega^{(\infty)}_2)(\bar{f}_n,f_n)
\nonumber\\
&=&\lim_{n\to\infty}2\langle E(f_n), |H|^{-1}F_{\beta}(|H|)E(f_n)\rangle,\nonumber
\end{eqnarray}
where $f_n$ is a sequence of real test-functions approaching a $\delta$-distribution at $x$. (The limit is well-defined, because all $\beta$-KMS states are Hadamard, so the difference $\omega^{(\beta)}_2-\omega^{(\infty)}_2$ is a smooth function on $M^{\times 2}$.)

We fix $0<\beta_0<\infty$. For $\beta_0\le\beta<\infty$ we can estimate
\[
F_{\beta_0}(k)-F_{\beta}(k)=F_{\beta_0}(k)\frac{e^{\beta k}}{e^{\beta k}-1}(1-e^{-(\beta-\beta_0)k})\ge0.
\]
It then follows from the spectral calculus that $\beta\to\omega^{(\beta)}_2(\Ws(x))$ is monotonically decreasing in $\beta\in(0,\infty)$. To prove continuity at $\beta_0$ we take $\beta\in[\frac12\beta_0,\infty)$ and we define
\[
G_k(\beta):=\frac{F_{\beta}(k)-F_{\beta_0}(k)}{F_{\frac14\beta_0}(k)}=\frac{e^{\frac14\beta_0k}-1}{e^{\beta k}-1}
-\frac{e^{\frac14\beta_0k}-1}{e^{\beta_0k}-1}
\]
which has $G_k(\beta_0)=0$ and
\[
\partial_{\beta}G_k(\beta)=\frac{-k}{1-e^{-\beta k}}\ \cdot\ \frac{e^{\frac14\beta_0 k}-1}{e^{\beta k}-1}<0.
\]
Note that $|\partial_{\beta}G_k(\beta)|$ is monotonically decreasing in $\beta$ and $\beta\ge\frac12\beta_0$, so
\[
|\partial_{\beta}G_k(\beta)|\le\left|\partial_{\beta}G_k\left(\frac12\beta_0\right)\right|
\le\frac{k}{1-e^{-\frac12\beta_0 k}}\ e^{-\frac14\beta_0 k}\le 2\beta_0^{-1}.
\]
This means that
\[
|F_{\beta}(k)-F_{\beta_0}(k)|\le2\beta_0^{-1}|\beta-\beta_0|F_{\frac14\beta_0}(k),
\]
and hence
\[
|\omega^{(\beta)}(\Ws(x))-\omega^{(\beta_0)}(\Ws(x))|\le C|\beta-\beta_0|
\]
for some $C\ge0$ depending on $\beta_0$ and $x$. This implies the continuity.

Finally, to determine the limit $\beta\to\infty$ we consider $0<\beta_0<\beta<\infty$ and we estimate
$F_{\beta}(k)\le F_{\beta_0}(k)\frac{\beta_0}{\beta}$. Hence we find
\[
0\le\omega^{(\beta)}_2(\Ws(x))-\omega^{(\infty)}_2(\Ws(x))
\le\frac{\beta_0}{\beta}\left(\omega^{(\beta_0)}_2(\Ws(x))-\omega^{(\infty)}_2(\Ws(x))\right),
\]
where the right-hand side vanishes as $\beta\to\infty$.
\end{proof*}

As long as the expectation values $\omega^{(\beta)}_2(\Ws(x))$ are non-negative, the local temperatures $T_{\omega^{(\beta)}}(x)$ of (\ref{Eqn_LT}) are well-defined, and they are continuous and monotonically decreasing with $\beta$, which is consistent with the definition of the global temperature (\ref{Eqn_GT}).

We will prove another useful result, for which we only need to assume that the theory has a ground state $\omega^{(\infty)}$ on $M$.
\begin{proposition}\label{Prop_minimum}
For every stationary two-point distribution $\omega_2$ and the ground state $\omega^{(\infty)}$ of (\ref{Eqn_ground}) we have that $\omega_2-\omega^{(\infty)}_2$ is of positive type.
\end{proposition}
For the $\beta$-KMS states this already follows from equations (\ref{Eqn_KMS},\ref{Eqn_ground}).
\begin{proof*}
We let $\omega$ be the quasi-free state with two-point distribution $\omega_2$. For any test-function $f\in C_0^{\infty}(M)$ we may denote the time-translated test-function by $\alpha_t(f)$, so that $\partial_t\alpha_t(f)|_{t=0}=\xi^a\nabla_af$. Let us fix any real-valued $f$ and define the function $w:\mathbb{R}\to\mathbb{C}$ by
\begin{equation}\label{Def_wt}
w(t-t'):=\omega_2(\alpha_{t'}(f),\alpha_t(f))=\langle\phi(f)\Omega_{\omega},e^{i(t-t')H}\phi(f)\Omega_{\omega}\rangle,
\end{equation}
where the last expression makes use of the GNS-representation of $\omega$, which has a Hamiltonian $H$ that implements the time evolution. Note that $w$ is continuous, bounded and of positive type, because $\omega_2$ is of positive type. This means that $w$ has a Fourier transform $\hat{w}$, which is a finite positive measure on $\mathbb{R}$, by Bochner's Theorem (\cite{Reed+} Theorem IX.9). Defining the even and odd parts of $w$ by $w_{\pm}(t):=\frac12(w(t)\pm w(-t))$ we see that $\hat{w}_-$ is an odd finite measure and $\hat{w}_+$ an even finite measure such that $\hat{w}_+\ge |\hat{w}_-|$.

We may define analogous functions $w^{(\infty)}$ for the ground state $\omega^{(\infty)}$. In this case we have additional properties:
$w^{(\infty)}$ is the boundary value of the holomorphic function
\begin{equation}\label{Eqn_wz}
w^{(\infty)}(z):=\langle\phi(f)\Omega_{\omega^{(\infty)}},e^{izH}\phi(f)\Omega_{\omega^{(\infty)}}\rangle,
\end{equation}
on the upper half plane in $\mathbb{C}$. This means that $\hat{w}^{(\infty)}$ has support in the half line $\mathbb{R}_{\ge0}$ (cf.~\cite{Hoermander1} Theorem 7.4.2 and 7.4.3). It follows that
\begin{equation}\label{Eqn_wground}
\hat{w}^{(\infty)}_+=|\hat{w}^{(\infty)}_-|
\end{equation}
on $\mathbb{R}\setminus\{0\}$. Note that $\hat{w}^{(\infty)}_+-|\hat{w}^{(\infty)}_-|=2\pi c\delta$ is a multiple of a $\delta$-distribution with some $c\ge0$. Because $w^{\infty}(t-t')-c$ is still of positive type we have
\begin{equation}\label{Eq_degeneracy}
0\le c\le\frac{1}{N^2}\sum_{j,k=0}^{N-1}w(j-k)=\left\|\frac{1}{N}\sum_{j=0}^{N-1}e^{ijH}\phi(f)\Omega_{\omega^{(\infty)}}\right\|^2
\end{equation}
for all $N\in\mathbb{N}$. In the quasi-free state $\omega^{(\infty)}$, $H$ preserves the one-particle Hilbert space and it is invertible there (cf.~equation (\ref{Eqn_ground})). The mean ergodic theorem (\cite{Reed+} Theorem II.11) then tells us that the right-hand side of (\ref{Eq_degeneracy}) vanishes in the limit $N\to\infty$. This means that $c=0$ and (\ref{Eqn_wground}) holds on all of $\mathbb{R}$.

Now note that $w^{(\infty)}_-=w_-$, because they only depend on the commutator distribution $\omega_2(x,y)-\omega_2(y,x)$. It then follows that $\hat{w}_+\ge|\hat{w}_-|=|\hat{w}^{(\infty)}_-|=\hat{w}^{(\infty)}_+$ and therefore $w_+-w^{(\infty)}_+$ is of positive type, hence
\begin{equation}\label{Eqn_statestimate}
\omega_2(f,f)-\omega^{(\infty)}_2(f,f)=w_+(0)-w^{(\infty)}_+(0)\ge0.
\end{equation}
Because $\omega_2-\omega_2^{\infty}$ is real-valued it follows that it is of positive type too.
\end{proof*}

When $\omega_2$ is Hadamard, $\omega_2-\omega_2^{(\infty)}$ is smooth and we conclude from proposition \ref{Prop_minimum} that
\begin{equation}\label{Eqn_Westimate}
\omega(\Ws(x))\ge\omega^{(\infty)}(\Ws(x))
\end{equation}
for all $x\in M$. Thus, if we want to prove lower bounds on $\omega(\Ws(x))$ for stationary Hadamard states, it suffices to consider the ground state. (See \cite{Pfenning+1998} for a related result.\footnote{I am grateful to Chris Fewster for directing my attention to this reference.})

\section{Existence of the local temperature}\label{Sec_existence}

To show that the local temperature in a general state $\omega$ is well-defined one needs to investigate whether $\omega(\Ws(x))\ge0$. In general this is not an easy problem, because we need to compare a global object (the state) with a local one (the Hadamard series). Moreover, it is typically false as we will see in section \ref{Sec_noT}. The main result of this section shows that stationary states do have a well-defined local temperature (\ref{Eqn_LT}) under suitable assumptions.
\begin{theorem}\label{Thm2}
Let $M=(\mathbb{R}\times\Sigma,g)$ be an ultra-static, globally hyperbolic space-time with compact Cauchy surface $\Sigma$ and non-trivial scalar curvature $R\ge0$. Let $O\subset\Sigma$ be an open neighbourhood where the Riemann curvature vanishes. Consider a massless free scalar field with scalar curvature coupling $\xi\in(0,\frac16)$. Then every stationary state has a well-defined local temperature at every $y\in O$.
\end{theorem}

Note that in ultra-static space-times the stationary observers are not rotating or accelerating, so there are no external forces on them. In these circumstances the weak energy condition on the classical background matter is equivalent to the strong and the dominant energy condition, which all state that the Ricci curvature $R_{ab}$ should be a positive operator. The positive curvature condition $R\ge0$ of theorem \ref{Thm2} is actually weaker than that.

In the following sections we will analyse the assumptions and provide a proof of theorem \ref{Thm2}. We will use a Wick rotation to express $\omega^{(\beta)}_2$ for any $\beta$-KMS state in terms of the fundamental solution $G$ of an elliptic operator on a compact, Riemannian manifold $\tilde{M}_{\beta}$. This leads to equation (\ref{Eqn_Wickomega2}) below, which allows us to recover $\omega^{(\beta)}(\Ws(y))$ from the behaviour of $G(x,\tilde{y})$, where $\tilde{y}\in\tilde{M}_{\beta}$ corresponds to $y\in M$ (see section \ref{SSec_Proof} for details). We then adapt an idea of Schoen \cite{Schoen1984} and show that we can perform a conformal transformation on $\hat{M}_{\tilde{y}}=\tilde{M}_{\beta}\setminus\{\tilde{y}\}$, which makes $\hat{M}_{\tilde{y}}$ asymptotically flat and which makes its ADM mass equal to $\omega^{(\beta)}(\Ws(y))$. Our strategy is to ensure that $\hat{M}_{\tilde{y}}$ verifies the assumptions of the positive mass theorem \cite{Schoen+1979,Schoen+1979_2}, which then entails $\omega^{(\beta)}(\Ws(y))\ge0$. We extend this estimate to all stationary states using the results of section \ref{Sec_monotonic}.

\subsection{The class of space-times}

Let $M$ be a (standard) static space-time, i.e.~$M=(\mathbb{R}\times\Sigma,g)$ with metric $g=-N^2dt^2+h$, where the smooth positive lapse function $N>0$ and the Riemannian metric $h$ on $\Sigma$ are independent of $t\in\mathbb{R}$. $M$ is globally hyperbolic if and only if $\Sigma$ is complete in the optical metric $N^{-2}h$ and in this case all the surfaces $\{t\}\times\Sigma$ are Cauchy \cite{Sanchez2005}. We call the space-time ultra-static when $N\equiv 1$. In this case $R$ equals the curvature of the Riemannian manifold $(\Sigma,h)$ by the Gauss-Codacci equations and the vanishing of the extrinsic curvature.

For theorem \ref{Thm2} we need compact manifolds $\Sigma$ with non-trivial $R\ge0$ which are flat in some open region $O$. In order to establish the existence of such manifolds we will make use of the following classical observation:
\begin{lemma}\label{Lem_Xi}
Let $\rho\in C_0^{\infty}(\mathbb{R}^3)$ be non-trivial and spherically symmetric with $\rho\ge0$ and $0\not\in\mathrm{supp}(\rho)$. Set $\mu:=\int\rho(x)dx$ and $\nu:=\int\frac{\rho(x)}{|x|}dx$ and define
\begin{equation}
U(x):=\int\frac{\rho(y)}{|x-y|}dy.\nonumber
\end{equation}
Then $-\Delta U=4\pi\rho$ with the Laplace operator $\Delta$. Furthermore, $U$ is spherically symmetric and monotonically decreasing with $|x|$, with $U(x)=\nu>0$ near $x=0$ and $U(x)=\frac{\mu}{|x|}>0$ for large $|x|$.
\end{lemma}
\begin{proof*}
$U$ is essentially the Newtonian gravitational potential for the mass distribution $\rho$, and the result is Newton's shell theorem
(\cite{Marion1965} chapter 5).
\end{proof*}

Choosing $U$ as in the lemma we endow $\mathbb{R}^3$ with the metric $g_{ij}:=U^4\delta_{ij}$, which has a non-trivial scalar curvature
\begin{equation}
R=-8U^{-5}\Delta U=32\pi U^{-5}\rho\ge0.
\end{equation}
For small $|x|$, $g_{ij}=\nu^4\delta_{ij}$ is flat. For large $|x|$ it is $\mu^4|x|^{-4}\delta_{ij}dx^idx^j$ and changing coordinates to $y:=\mu^2|x|^{-2}x$ it takes the Euclidean form $\delta_{ij}dy^idy^j$ on $|y|>0$. We may compactify $\mathbb{R}^3$ to the sphere $\mathbb{S}^3$ by adding the point $x=\infty$, i.e.~$y=0$, and we may extend the metric smoothly in $y=0$. This endows $\mathbb{S}^3$ with a Riemannian metric which has all the desired properties. Similar examples can be constructed in other dimensions, and one can construct further examples by performing small perturbations of the metric in the region where $R>0$ and by using gluing techniques \cite{Delay2011}.\footnote{I thank Stefan Hollands for pointing out these construction methods and for providing further concrete examples.}

\subsection{Wick rotation and $\beta$-KMS states}\label{SSec_Wick}

In the ultra-static case, $\beta$-KMS and ground states exist whenever $C_0^{\infty}(\Sigma)$ is in the domain of $\hat{K}^{-\frac12}$, where $\hat{K}$ is the Friedrichs extension of $K:=-\Delta_h+V$ on the Cauchy surface. To see this we refer to \cite{Sanders2013} Theorem 4.2 and 4.3, whose proofs still go through in this case. In the language of section \ref{Sec_monotonic} this verifies that $H$ is invertible and the domain of $|H|^{-1}$ contains all solutions $E(f)$ with $f\in C_0^{\infty}(M)$.

To verify the domain condition, we make use of the following classical result:
\begin{proposition}\label{Prop_Gpositive}
Consider a compact Riemannian manifold $(X,g)$ and a non-zero $V\in C^{\infty}(X)$ with $V\ge0$. Then the inverse $G$ of the Friedrichs extension of $K:=-\Delta_g+V$ has an integral kernel $G(x,y)$ with $G(x,y)>0$ when $x\not=y$.
\end{proposition}
\begin{proof*}
The conditions on $V$ ensure that $K>0$ has an invertible Friedrichs extension $\hat{K}>0$ with inverse $G>0$. For sufficiently large $N$, $(\hat{K}+I)^{-N}$ is a bounded operator with a continuous integral kernel, by elliptic regularity results (\cite{Hoermander1} Thm.~8.3.1). Because $X$ is compact this means that the operator $(\hat{K}+I)^{-N}$ is compact and has a discrete spectrum. It follows that $G$ also has a discrete spectrum and that it is bounded, which means that it has an integral kernel.

Now fix any $y\in X$ and note that $x\mapsto G(x,y)$ is real-valued (because $K$ has real coefficients) and smooth on $x\not=y$. Because $\lim_{x\to y}G(x,y)=+\infty$ we can find an open neighbourhood $O$ of $y$ such that $G(x,y)\ge1$ on $O\setminus\{y\}$. Let $\mu$ be the minimum of $G(.,y)$ over the compact set $U:=X\setminus O$. On the interior of $U$ we have $KG(.,y)=0$ and on the boundary $G(.,y)\ge 1$, so by the strong maximum principle (\cite{Evans} Sec.~6.4, Thm.~4) we see that $\mu>0$. Thus, $G(.,y)>0$ on $U$ and on $O\setminus\{y\}$.
\end{proof*}

For static space-times one can recover $\omega^{(\beta)}_2$ from a Wick rotation, as we will now recall (see e.g.~\cite{Sanders2013}). We first analytically continue $t$ to $z=t+i\tau$, where we choose $\tau$ to be $\beta$-periodic. We define the associated Riemannian manifold $\tilde{M}_{\beta}:=(\mathbb{S}^1\times\Sigma,\tilde{g})$ with metric $\tilde{g}=N^2d\tau^2+h$. Assuming $V$ is stationary, the elliptic operator analogous to the Klein-Gordon operator $K:=-\Box_g+V$ is $\tilde{K}:=-\Delta_{\tilde{g}}+V$, which we view as a symmetric operator on the domain of test-functions in $L^2(\tilde{M}_{\beta})$ (integration with the volume form induced by the metric). When $\Sigma$ is compact and $V\ge0$ is non-trivial, the Friedrichs extension of $\tilde{K}$ is invertible and its inverse defines the Euclidean Green's function $G^{(\beta)}$. This operator has an integral kernel $G^{(\beta)}(x,y)$ which is smooth away from the diagonal.

To recover the two-point distribution $\omega^{(\beta)}_2$ on $M$ one may analytically continue $G^{(\beta)}$ in $z$ and approach the real values $t$ from the correct side in the complex plane to ensure the Hadamard condition. Note in particular that for $t=t'=0$ and $x\not=x'\in\Sigma$ we have
\begin{eqnarray}\label{Eqn_Wickomega2}
\omega_2^{(\beta)}(0,x;0,x')=G^{(\beta)}(0,x;0,x').
\end{eqnarray}
This determines the $\beta$-KMS state on $t=t'=0$ away from the diagonal, and it provides enough information to determine the expectation values of the Wick square for all points on $t=0$. Because the time flow preserves the expectation values of the Wick square, this is all the information we will need.

An analogous procedure works for the ground state \cite{Wald1979}. As the simplest example of the Wick rotation we consider the Minkowski space $M_0=(\mathbb{R}^4,\eta)$ with inertial coordinates $(t,x^i)$. For a massless scalar field, $K_0\phi:=-\Box\phi=0$, the two-point distribution $\lambda_2$ of the Minkowski vacuum state $\lambda$ is given by
\begin{equation}\label{Eqn_Minkvac}
\lambda_2(t,x;t',x')=(2\pi)^{-2}\lim_{\tau\to0^+}(-(t-t'+i\tau)^2+|x-x'|^2)^{-1},
\end{equation}
where $|x-x'|$ denotes the Euclidean norm of the spatial coordinates. By definition of the Hadamard renormalisation scheme we have $\lambda(\Ws(x))\equiv 0$. On the Euclidean space $\tilde{M}_0=(\mathbb{R}^4,\delta)$, where $\delta$ is the Euclidean metric, we note that minus the Laplace operator $-\Delta$ is a strictly positive operator on the $L^2$-Hilbert space and its inverse $G_0$ has an integral kernel
\begin{equation}\label{Eqn_EuclMinkvac}
G_0(\tau,x;\tau',x')=(2\pi)^{-2}\left((\tau-\tau')^2+|x-x'|^2\right)^{-1}.
\end{equation}

\subsection{Proof of theorem \ref{Thm2}}\label{SSec_Proof}

Our proof adapts an idea of Schoen \cite{Schoen1984}:
\begin{proof*}
Fix $\beta>0$ and $y\in O\subset\Sigma$. We let $\tilde{y}=(0,y)$ be the corresponding point in the Riemannian manifold $\tilde{M}_{\beta}=(\mathbb{S}^1\times\Sigma,\tilde{g})$. Note that $x\mapsto G^{(\beta)}(x,\tilde{y})$ is a positive smooth function on $\tilde{M}_{\beta}\setminus\{\tilde{y}\}$ by proposition \ref{Prop_Gpositive}, so $\Omega(x):=4\pi^2 G^{(\beta)}(x,\tilde{y})$ is also a positive smooth function.

Let us choose Riemannian normal coordinates $x^i$ on $\Sigma$ centred on $y$, and extend them to Gaussian normal coordinates with $x^0:=\tau$.
Because the Riemann curvature vanishes near $y$ we then find that $\tilde{g}$ takes the Euclidean form on a neighbourhood of $\tilde{y}$ in $\tilde{M}_{\beta}$. Let $|x|$ denote the Euclidean metric in these coordinates and note that the Hadamard series for $G^{(\beta)}(x,\tilde{y})$ (with $m=0$ and $R=0$ near $\tilde{y}$) tells us that
\begin{equation}\label{Eqn_Hadamard}
\Omega(x)=4\pi^2G^{(\beta)}(x,\tilde{y})=|x|^{-2}+4\pi^2 w+O(|x|)
\end{equation}
near $\tilde{y}$, for some $w\in\mathbb{R}$. We note that $w=\omega^{(\beta)}(\Ws(y))$ is the expectation value of the Wick square in a thermal state. This can be seen by combining equation (\ref{Eqn_Wickomega2}) with the fact that for $x,y\in O\subset\Sigma$ the distribution $H_2^{(k)}$ takes the form $G_0(0,x;0,y)$ of the Euclidean Green's function in four dimensional Euclidean space (\ref{Eqn_EuclMinkvac}). Hence we find from (\ref{Eqn_Hadamard}) that
\begin{equation}
\omega^{(\beta)}(\Ws(y))=\lim_{x\to y}(G^{(\beta)}-G_0)(0,x;0,y)=w.
\end{equation}

Now consider the Riemannian manifold $\hat{M}_{\tilde{y}}:=(\tilde{M}_{\beta}\setminus\{\tilde{y}\},\hat{g})$ with $\hat{g}:=\Omega^2\tilde{g}$. This manifold is asymptotically flat, with $\tilde{y}$ the point at infinity \cite{Schoen1984,Geroch1972}. Indeed, in the coordinates $y^i:=|x|^{-2}x^i$ the metric $\hat{g}$ takes the form
\begin{equation}
\hat{g}=\Omega^2(|y|^{-2}y)|y|^{-4}\delta_{\mu\nu}dy^{\mu}dy^{\nu}=\left(1+8\pi^2 w|y|^{-2}+O(|y|^{-3})\right)\delta_{\mu\nu}dy^{\mu}dy^{\nu}.
\end{equation}
Note that the scalar curvature of $\tilde{M}_{\beta}$ equals the scalar curvature $R$ of $M$ (as a function on $\Sigma$, independent of $z=t+i\tau$). The scalar curvature $\hat{R}$ of $\hat{M}$ is then found to be
\begin{equation}
\hat{R}=\Omega^{-2}(R-6\Omega^{-1}\Delta_{\tilde{g}}\Omega)=(1-6\xi)\Omega^{-2}R,
\end{equation}
where we used $\tilde{K}\Omega=(-\Delta_{\tilde{g}}+\xi R)\Omega=0$ on $x\not=\tilde{y}$ in the last equality. The properties of $R$, $\Omega$ and $\xi$ together imply that $\hat{R}\ge0$ and $\hat{R}$ does not vanish identically. We may now apply the positive action theorem \cite{Schoen+1979_2} (which is the four-dimensional version of the positive mass theorem of general relativity \cite{Schoen+1979,Witten1981}) to conclude that $w\ge0$, i.e.~$\omega^{(\beta)}(\Ws(y))\ge0$. This holds for all $y\in O$ and all $0<\beta<\infty$. The claim for the ground state follows from taking the limit $\beta\to\infty$, using proposition \ref{Prop_monotonic}. For general stationary states the result then follows from proposition \ref{Prop_minimum}.
\end{proof*}

\section{Counter-examples}\label{Sec_noT}

In this section we indicate to what extent the assumptions of theorem \ref{Thm2} are necessary, by providing some examples of situations where the local temperature is not well-defined. All examples apply to points where $R=0$, so they are independent of any local renormalisation freedom for the Wick square.

\subsection{Non-stationary states}

It is well-known from the context of quantum inequalities that at any point $x\in M$ the expectation values $\omega(\Ws(x))$ are unbounded from below as $\omega$ ranges over all Hadamard states \cite{Fewster2005,Solveen2010}. This holds regardless of the renormalisation freedom (\ref{Eqn_freedom}), so there must be many states in which the expectation value is negative and the local temperature at $x$ ill-defined.

Quantum inequalities suggest to consider an average of $\omega(\Ws(x))$ along a time-like curve $\gamma$, weighted with a suitable positive weight function. Such an average can be bounded from below by a state-independent bound \cite{Fewster2005,Pfenning+1998,Fewster+2008}, and the analogous quantum inequality for the energy density can be interpreted in terms of the uncertainty relation of energy and time. Unfortunately, the lower bound is typically still negative. The largest bound is found if one smears over infinitely long times, in which case the bound can hardly be called local.

One may expect the situation to improve in a stationary space-time $M$ if we only consider stationary states. Let $\gamma_x:\mathbb{R}\to M$ be the unique curve which follows the stationary time flow and starts at $\gamma_x(0)=x$. Because the time-flow preserves the metric, it also preserves the distributions $H^{(k)}_2$ near the diagonal of $M\times M$. Consequently, for any stationary state $\omega$, the expectation value $\omega(\Ws(\gamma_x(\tau)))$ along the curve $\gamma_x$ is independent of $\tau$. Since the value $\omega(\Ws(x))$ is maintained ''forever'' along the entire curve $\gamma_x$, one might expect, by analogy with the time-energy uncertainty relations, that it must be non-negative. However, we will see in the next sections that this is not the case.

\subsection{Accelerated observers}

Another cause for the absence of local temperature is the acceleration of the stationary observers. The best known example is related to the Unruh effect. The Rindler space-time (given by the wedge $x^1>|x^0|$ in Minkowski space) is a static space-time for the Killing field of uniformly accelerated observers, and the restriction of the Minkowski vacuum state to the Rindler space-time is a $\frac{1}{2\pi}$-KMS state \cite{Unruh1976}. It is well understood that the difference in the local temperature which is assigned to the Minkowski vacuum by the inertial observers in Minkowski space-time and by the stationary observers in the Rindler wedge is due to the acceleration. In line with proposition \ref{Prop_monotonic} one may show that the Fulling vacuum in the Rindler space-time has a negative expectation value of the Wick square, which makes the local temperature (\ref{Eqn_LT}) in the ground state ill-defined (see e.g.~\cite{Buchholz+2013_3}).

We will briefly comment on the effects of acceleration beyond this well-known example. Although accelerated observers can occur in general stationary space-times, we will only discuss its effects in static space-times. (See also \cite{Eltzner2013} section 3.2 for related comments.)

Any static space-time $M=(\mathbb{R}\times\Sigma,g)$ is conformally equivalent to an ultra-static space-time $M'=(\mathbb{R}\times\Sigma,g')$ where $g=N^2g'$. $M$ is globally hyperbolic if and only if $M'$ is. In principle one can exploit this equivalence to study the expectation values of the Wick square in $\beta$-KMS states. Indeed, the Riemannian manifold associated with $M'$ is $\tilde{M}'_{\beta}=(\mathbb{S}^1\times\Sigma,\tilde{g}')$ with $\tilde{g}=N^2\tilde{g}'$. We note that the Laplace-Beltrami operators $-\Delta_{\tilde{g}'}$ and $-\Delta_{\tilde{g}}$ are related by
\begin{eqnarray}\label{Eqn_Kconf}
N^2(-\Delta_{\tilde{g}})N^{-2}=N^{-1}(-\Delta_{\tilde{g}'}+\frac16(R'-N^2R))N^{-1},
\end{eqnarray}
where the scalar curvatures $R$ and $R'$ of $M$ and $M'$ are related by
\begin{equation}
N^2R=R'-6N^{-1}(\Delta_{g'}N).
\end{equation}
The factors $N^2$ on the left-hand side of (\ref{Eqn_Kconf}) serve as unitary intertwiners between the $L^2$ spaces of $\tilde{M}_{\beta}$ and $\tilde{M}'_{\beta}$ (because we have four space-time dimensions). Consider the $\beta$-KMS states determined by the Euclidean Green's functions $G^{(\beta)}$ and $G'^{(\beta)}$ associated with the operators
\begin{eqnarray}
\tilde{K}=-\Delta_{\tilde{g}}+V,\qquad \tilde{K}'=-\Delta_{\tilde{g}'}+\frac16(R'-N^2R)+N^2V
\end{eqnarray}
for some non-trivial potential function $V\ge0$. Then (\ref{Eqn_Kconf}) implies
\begin{equation}\label{Eqn_Gconf}
G^{(\beta)}(x,y)=N^{-1}(x)G'^{(\beta)}(x,y)N^3(y).
\end{equation}
Using the distributions (or rather densities) $H_2^{(k)}$ and $(H'_2)^{(k)}$ in the two metrics $g$ and $g'$ one then finds
\begin{eqnarray}
\omega^{(\beta)}(\Ws(x))&=&N(x)^2\omega'^{(\beta)}(\Ws(x))\\
&&+\lim_{y\to x}(N(x)^{-1}(H'_2)^{(k)}(x,y)N(y)^3-H_2^{(k)}(x,y)).\nonumber
\end{eqnarray}
Instead of pursuing this analysis further we will only make two elementary comments here.

The change in the potential term from $V$ to $V':=\frac16(R'-N^2R)+N^2V$ enters non-locally when taking the inverse of $\tilde{K}'$. On the other hand, the distributions $H_2^{(k)}$ and $H_2^{(k)}$ are locally constructed from the metrics $g$ and $g'$, so they cannot cancel the non-local effects of this change. This can complicate the analysis of acceleration by conformal transformations, unless the field equation is conformally invariant, $V=\frac16R$.

In the simplest case $N\equiv c>0$ is a constant, which yields $c^2R=R'$ and $G^{(\beta)}=c^2G'^{(\beta)}$. The distributions $H_2^{(k)}$ also incur a factor $c^2$, which means that the $\beta$-KMS states $\omega^{(\beta)}$ and $\omega'^{(\beta)}$ satisfy
\begin{equation}
\omega^{(\beta)}(\Ws(x))=c^2\omega'^{(\beta)}(\Ws(x))
\end{equation}
and the local temperature in $M$ (if it exists) incurs a factor $c$. This rescaling of the local temperature can be interpreted as a mere change of units and it mirrors a similar rescaling for the global temperature. Indeed, one can rescale the Killing field by a constant factor $c$, which leads to a rescaled notion of global temperature. In some static space-times one can fix the scale factor by imposing additional physically motivated conditions, e.g.~by requiring that $\xi^a\xi_a=-1$ near space-like infinity in asymptotically flat space-times. Note that ultra-static space-times in particular do not suffer from the same freedom in the global choice of scale.

\subsection{Violation of energy conditions}\label{SSec_Rneg}

In this section we will construct new examples of ground states in ultra-static space-times with ill-defined local temperatures. Here the problem cannot be attributed to accelerated observers or the local renormalisation freedom of the Wick square. Instead we claim that it is due to a violation of the curvature condition $R\ge0$.

Before we give our examples we will present a useful comparison result for elliptic operators on the Euclidean side.
\begin{proposition}\label{Prop_ellipticcomparison}
Let $(X,g)$ be a complete Riemannian manifold and $V\in C^{\infty}(X)$ with $V\ge 0$. Define the operator $K_1:=-\Delta_g+V$ on $C_0^{\infty}(X)\subset L^2(X)$ and assume that the Friedrichs extension of $K_1$ has an inverse $G_1$ given by an integral kernel $G_1(x,y)$. Finally, let $K_2$ be another elliptic second order differential operator on $X$ such that $K_2\ge K_1$.

Then the Friedrichs extension of $K_2$ has an inverse $G_2$ satisfying $0<G_2\le G_1$ and given by an integral kernel $G_2(x,y)$. When $K_1\equiv K_2$ on an open neighbourhood $O\subset X$, then $G_1-G_2$ is smooth on $O\times O$, and $(G_1-G_2)(x,x)\ge 0$ for all $x\in O$. If in addition $G_1(x,y)>0$ on $x\not=y$ in $X^{\times 2}$, if $K_2-K_1\ge \kappa\ge0$ for some $\kappa\in C_0^{\infty}(X)$ and if $(G_1-G_2)(x,x)=0$ for all $x$ in some open subset $O'\subset O$, then $G_1=G_2$ and hence $K_1\equiv K_2$ everywhere.
\end{proposition}
\begin{proof*}
The estimate $0<G_2\le G_1$ is well known, and the existence of the integral kernel $G_1(x,y)$ then implies the existence of $G_2(x,y)$. On $O\times O$ the distribution $(G_1-G_2)(x,y)$ satisfies
\begin{equation}
K_{2,x}(G_1-G_2)(x,y)=0=K_{2,y}(G_1-G_2)(x,y),
\end{equation}
where the subscript on $K_2$ indicates the variable on which it acts. Note that $K_{2,x}+K_{2,y}$ is elliptic on $X^{\times 2}$, so by standard elliptic regularity results (\cite{Hoermander1} Thm.~8.3.1) this implies that $G_1-G_2$ is smooth on $O\times O$. Because $G_1-G_2$ is of positive type we find $(G_1-G_2)(x,x)\ge0$ for $x\in O$ (cf.~\cite{Reed+} Sec.IX.2).

Now assume that $G_1(x,y)>0$ when $x\not=y$, that $K_2-K_1\ge \kappa\ge0 $ for some $\kappa\in C_0^{\infty}(X)$ and that there is an open subset $O'\subset O$ with $(G_1-G_2)(x,x)=0$ for all $x\in O'$. We will assume in addition that $\kappa\not=0$ and derive a contradiction. Because $\kappa\equiv0$ on $O$ we may choose a compact ball $B\subset X$ containing a neighbourhood where $\kappa\not=0$ in its interior and such that $B\cap O'=\emptyset$ and the complement $B^c:=X\setminus B$ is connected. Then we may choose a $W\in C_0^{\infty}(X)$ supported in $B$ and such that $0\le W\le \kappa$ and $W\not\equiv0$. Set $K_3:=K_1+W$ and let $G_3$ be the inverse of its Friedrichs extension, so that $G_1\ge G_3\ge G_2>0$. It follows from our assumptions that $(G_1-G_3)(x,x)$ vanishes for all $x\in O'$, but $G_1\not= G_3$. We will now derive a contradiction from these statements.

For all positive test-functions $f,f'$ on $X$ we have
\begin{eqnarray}
0&\le& (G_1-G_3)(f\pm f',f\pm f')\nonumber\\
&=&(G_1-G_3)(f,f)\pm2(G_1-G_3)(f',f)+(G_1-G_3)(f',f')
\end{eqnarray}
by symmetry of the $G_i$. The function $x\mapsto(G_1-G_3)(x,f)$ is smooth (by elliptic regularity) and $G_1-G_3$ is smooth on $B^c\times B^c$ by the first paragraph of this proof. We can therefore let $f'$ approximate a positive multiple $\lambda$ times a delta-distribution at some $x\in O'$ to find that $0\le (G_1-G_3)(f,f)\pm2\lambda(G_1-G_3)(x,f)$. Taking $\lambda\to\infty$ we see that we must have $(G_1-G_3)(x,f)= 0$ for all $x\in O'$. Now the smooth function $H_f(x):=(G_1-G_3)(x,f)$ vanishes on $O'$ and it satisfies $K_3H_f(x)=W(x)G_1(x,f)$, which vanishes on $B^c$. Using a unique continuation result (\cite{Hoermander3} Thm.~17.2.6) we conclude that $H_f$ must vanish on the connected set $B^c$. In particular, $H_f$ attains a non-positive minimum at an interior point of $X$. On the other hand, our positivity assumptions imply that $W(x)G_1(x,f)\ge0$, so by the strong maximum principle (\cite{Evans} Sec.~6.4, Thm.~4) we conclude that $H_f$ is constant. Thus $(G_1-G_3)(.,f)=H_f\equiv 0$ for all positive $f$. Because every test-function can be written as a linear combination of positive test-functions we find $(G_1-G_3)(.,f)=H_f\equiv 0$ for all $f$, which yields the desired contradiction that $G_1-G_3=0$.
\end{proof*}

Let us consider as an example the equation $K\phi:=(-\Box+V)\phi=0$ in Minkowski space $M_0$, where $V\ge0$ is independent of the inertial time coordinate $t$. We can obtain $\omega_2^{(\infty)}$ from a Wick rotation, starting with the elliptic operator $\tilde{K}:=-\Delta+V$. Note that $\tilde{K}\ge-\Delta>0$ has an inverse $G$ with $0<G\le G_0$. Now suppose that $V$ vanishes on an open neighbourhood $O$ in the $t=0$ plane. Using proposition \ref{Prop_ellipticcomparison} we then find that
\begin{equation}\label{Eqn_WicksquareWickrotation}
\omega^{(\infty)}(\Ws(x))=\lim_{y\to x}(\omega^{(\infty)}_2-\lambda_2)(y,x)=(G-G_0)(x,x)\le 0
\end{equation}
for all $x\in O$. Moreover, if $V$ does not vanish identically, then the inequality must be strict on a dense subset of $O$.

We now want to improve this example by showing that we can also have $V\equiv 0$ (i.e.~$\xi=0$ as well as $m=0$) in suitably chosen space-times. Let $M=(\mathbb{R}^4,g)$ with $g=-dt^2+h$ and $h=\Omega^2\delta$, where $\delta$ is the spatial Euclidean metric and $\Omega\in C^{\infty}(\mathbb{R}^3)$ has $\Omega>0$. Note that the scalar curvature of $M$ equals that of the Cauchy surface in the metric $h$, which  is given by
\begin{equation}\label{Eqn_R}
R=-4\Omega^{-2}\Delta(\ln\Omega)-2\Omega^{-2}\delta^{ij}(\partial_i\ln\Omega)(\partial_j\ln\Omega).
\end{equation}
We now make the more specific choice $\Omega(x):=e^{\nu-U(x)}$ with $\nu$ and $U$ as in lemma \ref{Lem_Xi}. It follows that $\Omega\ge1$ with equality near $x=0$, and $\Omega$ is independent of $t$. From $\Delta(\ln\Omega)(x)=4\pi\rho(x)\ge0$ and equation (\ref{Eqn_R}) we find that
\begin{equation}
R=-4\Omega^{-2}\Delta(\ln\Omega)-2\Omega^{-2}\delta^{ij}(\partial_i\ln\Omega)(\partial_j\ln\Omega)\le0\nonumber
\end{equation}
is non-trivial when $\rho\ge0$ is non-trivial.

The space-time $M$ is static and globally hyperbolic, because $\Omega\ge1$ \cite{Sanchez2005}. We consider its Riemannian counterpart $\tilde{M}=(\mathbb{R}^4,\tilde{g})$ with $\tilde{g}=d\tau^2+h$. The Laplace-Beltrami operator $-\Delta_{\tilde{g}}$ has an invertible Friedrichs extension in $L^2(\mathbb{R}^4,d\mathrm{vol}_{\tilde{g}})$, where the natural volume form is $d\mathrm{vol}_{\tilde{g}}=\Omega^3d^4x$. It is convenient to go over to $L^2(\mathbb{R}^4,dx^4)$, using multiplication by $\Omega^{\frac32}$ as a unitary intertwiner between the $L^2$ spaces. Accordingly we consider the operator
\begin{eqnarray}\label{Eqn_conformalequality}
\Omega^{\frac32}(-\Delta_{\tilde{g}})\Omega^{-\frac32}&=&-\partial_{\tau}^2+\Omega^{\frac32}(-\Delta_h)\Omega^{-\frac32}\\
&=&-\partial_{\tau}^2+\Omega^{-1}(-\Delta)\Omega^{-1}-\frac18R.\nonumber
\end{eqnarray}
This defines a symmetric operator on the test-functions. Because $\Omega^{-1}\le1$ and $\Omega$ is independent of $t$ we find from $R\le0$
that
\begin{equation}\label{InEq_conformalcomparison}
\Omega^{-1}(-\partial_{\tau}^2-\Delta)\Omega^{-1}\le\Omega^{\frac32}(-\Delta_{\tilde{g}})\Omega^{-\frac32}+\frac18R
\le \Omega^{\frac32}(-\Delta_{\tilde{g}})\Omega^{-\frac32}.
\end{equation}
It follows that the Friedrichs extension of $\Omega^{\frac32}(-\Delta_{\tilde{g}})\Omega^{-\frac32}$ has an inverse $G$ with $G\le\Omega G_0\Omega$, with $G_0$ as defined in (\ref{Eqn_EuclMinkvac}). This implies in particular that $G$ is given by an integral kernel $G(x,y)$. We denote the corresponding ground state by $\omega^{(\infty)}$ as usual.

On a neighbourhood $O$ of $x=0$ we have $\Omega\equiv 1$ and hence $g=\eta$, so to compute $\omega^{(\infty)}(\Ws)$ on $O$ we only need to take the coinciding points limit in $G-G_0=\Omega^{-1}G\Omega^{-1}-G_0$ on $O\times O$. We conclude from proposition \ref{Prop_ellipticcomparison} that $G-\Omega G_0\Omega$ has a smooth integral kernel on $O\times O$, and unless $R$ vanish identically, there are points $x\in O$ such that
\begin{equation}
\omega^{(\infty)}(\Ws(x))=(G-\Omega G_0\Omega)(x,x)<0.
\end{equation}
Thus we have established the existence of ground states in ultra-static space-times with negative expectation values of the Wick square, even in a region where $R=0$.

Using $R\le0$ our counter-example can easily be extended to scalar fields with non-minimal scalar curvature coupling $\xi<\frac18$. In this case the right-hand side of (\ref{InEq_conformalcomparison}) contains the operator $-\Delta_{\tilde{g}}+\xi R$ and the non-positive term $(\frac18-\xi)R\le0$. In the limiting case $\xi=\frac18$, which corresponds to conformal coupling in the three spatial dimensions, the Wick-square might vanish throughout the region $O$.


\subsection{A further existence result for local temperature}\label{Sec_LT}

Adapting the techniques of section \ref{SSec_Rneg} one may also prove a positive result for space-times with a non-compact Cauchy surface:
\begin{theorem}\label{Thm1}
Let $M=(\mathbb{R}^4,g)$ with $g=-dt^2+h$ and $h=\Omega^2\delta$ conformally related to the spatial Euclidean metric. Assume that $\Omega$ has the following properties: $\Omega$ is independent of $t$, $\Omega^{-1}\ge1$ is bounded, $\Omega\equiv1$ in an open neighbourhood $O$, and $R$ as given in (\ref{Eqn_R}) has $R\ge0$. Consider a massless free scalar field with scalar curvature coupling $\xi\in[0,\frac18]$. Then every stationary state has a well-defined temperature at every $x\in O$.
\end{theorem}
\begin{proof*}
Note that $M$ is globally hyperbolic, because $\Omega^{-1}$ is bounded. We reconsider the construction of our examples in section \ref{SSec_Rneg} and consider the operator $\Omega^{\frac32}(-\Delta_{\tilde{g}}+\xi R)\Omega^{-\frac32}$ on the Euclidean $\mathbb{R}^4$. For the spatial part we see from $R\ge0$ and from $\Omega\ge c$ with some $c>0$ that
\[
\Omega^{\frac32}(-\Delta_h+\xi R)\Omega^{-\frac32}\ge -\Omega^{-\frac32}\partial_i\delta^{ij}\Omega\partial_j\Omega^{-\frac32}
\ge c\Omega^{-\frac32}(-\Delta)\Omega^{-\frac32}.
\]
The right-hand side has an invertible Friedrichs extension on $\mathbb{R}^3$, with an inverse given by the integral kernel $c^{-1}\Omega^{\frac32}(x)G_0(x,y)\Omega^{\frac32}(y)$. It follows that the Friedrichs extension of the left-hand side is also invertible and that the inverse is given by an integral kernel. This in turn ensures the existence of a ground state. To prove the theorem it suffices to consider the ground state (see proposition \ref{Prop_minimum}). In analogy to equation (\ref{InEq_conformalcomparison}) we now find
\[
\Omega^{-1}(-\partial_{\tau}^2-\Delta)\Omega^{-1}\ge \Omega^{\frac32}(-\Delta_{\tilde{g}}+\xi R)\Omega^{-\frac32},
\]
using $\Omega^{-2}\ge1$ and $R\ge 0$. The result on $O$ then follows from proposition \ref{Prop_ellipticcomparison}, by the same argument as in section \ref{SSec_Rneg}.
\end{proof*}

Functions $\Omega$ satisfying the assumptions needed for theorem \ref{Thm1} can be constructed again from lemma \ref{Lem_Xi}. E.g., $\Omega(x):=\frac12+\frac{1}{2\nu}U(x)$ satisfies $1\ge\Omega(x)\ge\frac12$ with $\Omega\equiv1$ near $x=0$. From $\Delta\Omega(x)=-\frac{2\pi}{\nu}\rho(x)$ and a rewritten version of equation (\ref{Eqn_R}) we find that
\begin{equation}
R=-4\Omega^{-3}\Delta\Omega+2\Omega^{-4}\delta^{ij}(\partial_i\Omega)(\partial_j\Omega)\ge0
\end{equation}
is non-trivial when $\rho$ is non-trivial.

\section{Discussion}\label{Sec_discussion}

We have established some basic properties of expectation values of the Wick square of free scalar fields, to verify its use as a local thermometer. This includes the continuity and monotonicity of the map $\beta\mapsto\omega^{(\beta)}(\Ws(x))$ in general stationary space-times, and the fact that the ground state provides a lower bound for $\omega(\Ws(x))$ in all stationary states. However, the main focus of attention was on the question whether the expectation values for massless fields are non-negative, which is needed to define the local temperature.

Whereas the global temperature of a ground state is always zero (by definition), we have seen that a local temperature may not even be well-defined. One known cause is the acceleration of the stationary observers, which in general can have a complicated effect on the expectation values $\omega^{(\infty)}(\Ws(x))$. Another cause which we have identified is the violation of the classical energy conditions by the background metric. We have shown this by constructing explicit examples and we ensured that $x$ lies in a flat region of an ultra-static space-time, to make it clear that the negative expectation values are not due to acceleration or other local physical effects. On the positive side we have shown that $\omega(\Ws(x))\ge0$ for all stationary states in the absence of acceleration or rotation, when $R\ge0$ and when a few other technical assumptions are satisfied. The main theorem \ref{Thm2} assumes compact Cauchy surfaces, but we obtained a similar result for certain space-times with non-compact Cauchy surfaces as well (theorem \ref{Thm1}). Our results suggest that the global and local notions of temperature provide qualitatively similar information, as long as certain physically relevant restrictions are imposed.

There are various open questions worthy of further consideration. Although theorem \ref{Thm2} attains a fair level of generality, one may wonder if it extends to other potential functions (c.f.~\cite{Hermann+2014}), points where the space-time is not flat (possibly using the formula of \cite{Lynch+2016} instead of equation (\ref{Eqn_LT})), or to space-times with non-compact Cauchy surfaces (possibly with additional assumptions like asymptotic flatness). In addition to the Wick square, one may also consider other local observables, as suggested by \cite{Buchholz+2002_2}. Furthermore, for massive theories the relation between $\beta$ and $\omega^{(\beta)}(\Ws(x))$ becomes more complicated also in Minkowski space. To define a local temperature for massive theories one should therefore investigate different (non-zero) lower bounds on the expectation value of the Wick square. The renormalisation freedom in the Wick square, which involves the scalar curvature and the mass, should then be addressed as well. We hope to investigate some of these questions in the future.

\section*{Acknowledgements}

I would like to thank Chris Fewster for his suggestion that ''There ought to be some statement about QFT in the positive mass theorem''. I
am also grateful to him and to Stefan Hollands, Rainer Verch and Michael Gransee for their encouragement and helpful comments, and to Peter Taylor for bringing a further reference to my attention. Finally I would like to thank the anonymous referees for their careful reading of the manuscript and their comments.

\end{document}